\documentstyle[prc,preprint,tighten,aps]{revtex}

\preprint{\vbox{. \hfill TRI-PP-96-03, UM-P-96/23, RCHEP 96/2}}

\title{$\bf \Delta I = 1$ axial-vector mixing and charge symmetry breaking}

\author{S.\ A.\ Coon\thanks{Electronic-mail address: coon@nmsu.edu}
\thanks{Permanent address: Physics Department, New Mexico State University,
         Las Cruces, NM  88003, USA}}
\address{TRIUMF, 4004 Wesbrook Mall, Vancouver, B.\ C.\ Canada  V6T 2A3
	 \protect\\ and
 	 School of Physics, Research Center for High Energy Physics,
         University of Melbourne, Parkville, Victoria, Australia, 3052}
\author{B.\ H.\ J.\ McKellar\thanks{Electronic-mail address:
        mckellar@physics.unimelb.edu.au}}
\address{School of Physics, Research Center for High Energy Physics,
         University of Melbourne, Parkville, Victoria, Australia, 3052}
\author{V.\ G.\ J.\ Stoks\thanks{Electronic-mail
        address: stoks@alph02.triumf.ca}}
\address{TRIUMF, 4004 Wesbrook Mall, Vancouver, B.\ C.\ Canada  V6T 2A3}

\begin{document}

\maketitle

\begin{abstract}
Phenomenological Lagrangians that exhibit (broken) chiral symmetry as
well as isospin violation suggest short-range charge symmetry breaking
(CSB) nucleon-nucleon potentials with a $\mbox{\boldmath $\sigma$}_1
\!\cdot\!\mbox{\boldmath $\sigma$}_2$
structure. This structure could be realized
by the mixing of axial-vector ($1^+$) mesons in a single-meson exchange
picture. The Coleman-Glashow scheme for $\Delta I_{z}=1$ charge symmetry
breaking applied to meson and baryon $SU(2)$ mass splittings suggests a
universal scale.  This scale can be extended to $\Delta I=1$ nonstrange
CSB transitions $\langle a_1^\circ|H_{em}|f_1\rangle$ of size
$-0.005$ GeV$^2$.  The resulting nucleon-nucleon axial-vector meson
exchange CSB potential then predicts $\Delta I=1$ effects which are
small.
\end{abstract}
\vskip20pt
\noindent PACS numbers: 14.40.Cs, 13.75.Cs, 21.30.-x
\vskip20pt


\baselineskip 0.6cm               

\newpage

\section{INTRODUCTION}
A recent assessment of isospin violation in the nucleon-nucleon force
was made with the aid of phenomenological Lagrangians that exhibit
(broken) chiral symmetry as well as isospin breaking~\cite{vkfg}.  The
charge symmetry breaking (CSB) nucleon-nucleon ($N\!N$) forces arise from the
Lagrangian with the aid of dimensional power counting~\cite{Bira},
so that the Lagrangian suggests specific pion-exchange $N\!N$ force
mechanisms.  In addition,
four-nucleon terms in the Lagrangian lead to $N\!N$  contact forces with a
given spin and isospin structure.  The
undetermined (dimensionless) coefficients of these forces are expected
to be ``natural," i.e. of order ${\cal O}$(1) when written in the
appropriate scale~\cite{Weinberg,GM}. Therefore, a part of this
assessment~\cite{vkfg} included a comparison of these generic CSB forces
with specific models such as the standard $\pi$--$\eta$ and $\rho$--$\omega$
mixing models~\cite{PSC,PSC77,CSeta,CB} which successfully describe nuclear
data~\cite{vanoers}.  The  strength of the meson mixing forces and of
their pseudoscalar and vector-meson mixing matrix elements was
found to be ``natural" by this criterion~\cite{vkfg}.  The latter
meson-mixing matrix elements have been found to have a universal
scale~\cite{CS95}, obtained with the aid of the Coleman-Glashow picture
that charge symmetry-violating processes are dominated by CSB tadpole
diagrams~\cite{CG,Coleman}. Then the generic analysis of charge
symmetry breaking in the
$N\!N$ force appears consistent with these earlier specific mechanisms.
However, the generic Lagrangian of Refs.~\cite{vkfg,Bira} includes a
term of the form
\[
\gamma_{\sigma}(\bar{N}\frac{\tau_3}{2}\mbox{\boldmath $\sigma$}N)\!\cdot\!
               (\bar{N}\mbox{\boldmath $\sigma$}N)\ ,
\]
which leads to a class III (i.e., proportional to ($\tau_3(1) +
\tau_3(2)$)~\cite{vanoers}) CSB $N\!N$ force, apparently
not envisaged before.  In this paper, we study the suggestion that the
simple $t$-channel exchange of axial-vector
mesons ($1^+$) might be the dominant contribution to this proposed
short-range CSB $N\!N$ force.

The $N\!N$ contact force of the chiral Lagrangian represents the sum of
all $t$-channel exchanges (multi-pion resonances, uncorrelated pion loop
diagrams, simultaneous pion-photon loop diagrams, etc.)
 with a given spin and
isospin structure. However, the effective
chiral Lagrangian analysis can say nothing about the relative size of the
individual contributions to the parameter $\gamma_{\sigma}$ (nor to the
other short-range term $\gamma_s(\bar{N}\frac{\tau_3}{2}
N)\!\cdot\! (\bar{N}N)\ $) and one must turn to modeling. (In any case,
models are essential if one is to use the resulting CSB potentials in nuclear
calculations.)
  In the well-studied mesonic sector
it has been found that the ten phenomenological  parameters
$L_1\ldots L_{10}$ of the chiral Lagrangian (to one
loop) are saturated by  the exchange of the low-lying multi-pion resonances
($\rho$, $\omega$, $a_1$, $\eta_8$)  to the extent that
``there is no indication for the presence of any other contribution in
addition to the meson resonances"~\cite{Ecker}.  In the nucleonic
sector, the analogous parameters of the Lagrangian which generates
one-pion-range potentials have now been established, but
 the  short-range CSB $N\!N$ parameters $\gamma_s$ and
$\gamma_{\sigma}$ have not been determined
phenomenologically~\cite{vkfg}.  All that
one can say, at this stage, is that
models based on isospin mixing of meson resonances are
consistent with the naturalness criterion~\cite{vkfg} and with the
nuclear data~\cite{vanoers,CS95} when
the on-mass-shell mixing matrix element of the Coleman-Glashow picture
is employed. Still, these results in the mesonic and nucleonic sectors
suggest dominance of $\gamma_{\sigma}$ by
a simple $t$-channel (axial-vector) meson resonance mixing mechanism.
To study this hypothesis we employ the
Coleman-Glashow mixing
model which (unlike other mixing models) is immediately extended
to the axial-vector mesonic resonances $a_1$ and $f_1$, where there is
no experimental data on isospin mixing.

In this Coleman-Glashow picture, the meson-mixing matrix element
$\langle a_1^\circ|H_{em}|f_1\rangle$ is determined by
the dominant single-quark operator
$H^{(3)} = {\textstyle \frac{1}{2}}(m_u - m_d)\bar{q}\lambda_3 q$,
established in Ref.~\cite{CS95} from the electromagnetic mass
differences in the pseudoscalar mesons, vector mesons, baryon octet, and
baryon decuplet.  When extended to the off-diagonal $\Delta I = 1$
transitions $\langle\rho^\circ|H_{em}|\omega\rangle$ and
$\langle\pi^\circ|H_{em}|\eta_{NS}\rangle$, this gives the value
$-0.005$ GeV$^2$ for the mixing matrix elements, independent of the
particular mesons concerned.  Indeed, for this one-body operator, it is
to be expected that one obtains the same numerical value connecting any
$I=1$ state with an $I=0$ nonstrange state of the same spin and parity.
Thus, for this first estimate of this novel source of charge
symmetry breaking we assume $\langle a_1^\circ|H_{em}|f_{1\,NS}\rangle
\approx -0.005$ GeV$^2$, and refer to a discussion of delta meson
tadpole graphs in Ref.~\cite{CS95} for more details.

{}From this single-quark operator picture, or the equivalent
tree-level tadpole picture, the electromagnetic transition $\langle
a_1^\circ|H_{em}|f_{1\,NS}\rangle$ could have no dependence on the
four-momentum squared of the hadrons. Therefore, the analogous
theoretical predictions for the pseudoscalar and vector mesons are
compared with measured electromagnetic mass splittings and with
(on-mass-shell) CSB
violating transitions $\eta\rightarrow 3\pi$ and $\omega \rightarrow
2\pi$.  The comparison is excellent~\cite{CS95} and encourages us to
extend this scheme to the axial-vector mesons.
We emphasize that single-quark operator or tadpole-generated
mixings of the Coleman-Glashow scheme are not of the two-body current
$\times$ current type with conserved currents for which
the vector-meson mixing amplitude must identically
vanish at four-momentum squared equal to zero~\cite{OZ1}.
Furthermore, the claims based on various grounds~\cite{OZ1,qdep} that the
on-mass-shell $\rho$--$\omega$ mixing matrix element is greatly suppressed
would force an unnaturally small coefficient $\gamma_s$ for the
resulting CSB term of the $N\!N$ force if the $\rho$--$\omega$ mixing
mechanism saturates that term~\cite{vkfg}. By the criteria of
Refs.~\cite{vkfg,Bira,Weinberg,GM}, such an unnaturally small coefficient
would presuppose a symmetry to preserve its small value.  Such a symmetry
has not been identified, nor has an alternate mechanism been found which
could both restore naturalness to $\gamma_s$ and describe all the
nuclear data.  For these reasons, we will use the
on-mass-shell $a_1$--$f_{1\,NS}$ mixing matrix element obtained from the
Coleman-Glashow scheme, rather than follow the
suggestions~\cite{OZ1,qdep} of a suppression of particle mixing at the
off-mass-shell kinematics of an $N\!N$ force diagram.  We return to a
discussion of the saturation argument after presenting our results.

In the following, we will derive the CSB potential due to
$a_{1}$--$f_{1}$
mixing, give two estimates for the coupling constants involved, and
present the results for the $N\!N$ singlet scattering lengths and
for the ${^{3}}$H--${^{3}}$He binding-energy difference.  As we will
show, the $f_1$ meson is primarily nonstrange ($NS$) so we drop the
distinction between $f_{1\,NS}$ and $f_1$ at this point.

\section{CSB POTENTIAL DUE TO $\protect
         \lowercase{a_1}$--$\protect
         \lowercase{f_1}$ MIXING}
We begin with the total interaction Hamiltonian~\cite{BD,footnote}
\begin{equation}
   H_I(a_1N\!N)={\textstyle\frac{1}{2}} g_{a_1N\!N}
        \bar{N}\mbox{\boldmath $\tau$}\!\cdot\!{\bf A}^{\mu}
        \gamma_{\mu}\gamma_5 N
   \; +\; {\textstyle\frac{1}{2}} g_{f_1N\!N}
        \bar{N} f^{\mu} \gamma_{\mu}\gamma_5 N
   \; +\; X A^{\mu}f_{\mu}\ ,         \label{eqno1}
\end{equation}
which defines the $AN\!N$ coupling constants, and where $X$ simulates the
electromagnetic $a_1$--$f_1$ transition. We have neglected the
tensor couplings from the beginning.
Experience with the vector mesons, where
the momentum-dependent tensor coupling of the $\rho$ is especially large,
shows that the class III CSB results are merely reduced by 10-20\% by
this neglect~\cite{CB,footnote2}.

In the actual $a^\circ_1$--$f_1$ $N\!N$ force diagram,
the Feynman rules give ($S_{fi} = I_{fi} - iT_{fi}(2\pi)^4\delta(P_{fi}))$
\begin{eqnarray}
   T^{a_1f_1}_{N\!N} &=& +{\textstyle\frac{1}{4}} g_{a_1N\!N}g_{f_1N\!N}
       \frac{X}{(m^2_{a_1}-t)(m^2_{f_1}-t)} \nonumber  \\
   & & \mbox{} \times
       \bar{u}(p_{1f})\tau_3\gamma_{\mu}\gamma_5u(p_{1i})
       \left(g^{\mu\nu}-\frac{q^{\mu}q^{\nu}}{m^2_{a_1}}\right)\,
       g_{\nu\sigma}\,
       \left(g^{\sigma\tau}-\frac{q^{\sigma}q^{\tau}}{m^2_{f_1}}\right)
       \bar{u}(p_{2f})\gamma_{\tau}\gamma_5 u(p_{2i})    \nonumber\\
   & & \mbox{} + (1\longleftrightarrow 2)\ ,       \label{eqno2}
\end{eqnarray}
with $q=(p_{1i}-p_{1f})=(p_{2f}-p_{2i})$, $t \equiv q^2$, and
we have taken a narrow $a_1$ width to simplify the calculation.

The term $X$ in the interaction Hamiltonian is related to the
electromagnetic $a_1$--$f_1$ transition as explained in
Ref.~\cite{PSC77}.  That is, the scalar $\langle
a_1^\circ|H_{em}|f_1\rangle$ will not change the helicity of the massive
spin-1 meson. So we need consider only the specific helicity states in
Eqs.~(\ref{eqno1}) and (\ref{eqno2}),
\begin{equation}
   \langle a_1^{(\lambda)} |H_I| f_1^{(\lambda)}\rangle
   = X \langle a_1^{(\lambda)} |A^{\mu}f_{\mu}| f_1^{(\lambda)}\rangle
   = X \varepsilon^{*(\lambda)}_{\mu}\varepsilon^{\mu (\lambda)}
   =-X\ ,                   \label{eqno3}
\end{equation}
where we have used the identity
$\varepsilon^{*(\lambda^{\prime})}_{\mu}\varepsilon^{\mu (\lambda)} =
-\delta^{\lambda^{\prime} \lambda}$.
Since the left-hand side of Eq.~(\ref{eqno3})
is the electromagnetic $a_1$--$f_1$ transition, it is clear that the
contact term replacing $X$ in the Feynman
graph for the $N\!N$ force must be
$-\langle a_1^\circ|H_{em}|f_1\rangle$.

Because the mass of the $a_1$ is poorly
known~\cite{PDG} and, in any event, nearly degenerate with that of the
$f_1(1285)$, we now take the limit $m_{a_1}=m_{f_1}$ arguing that the
corrections ${\cal O}(m_{a_1}-m_{f_1}$) are surely smaller than any other
uncertainty in this problem. With this approximation, we specialize to the
$S$-wave, make a
nonrelativistic reduction of Eq.~(\ref{eqno2}) to ${\cal O}$$({\bf q}^2)$,
where ${\bf q}$ is the momentum transferred to the axial-vector meson,
${\bf q}={\bf p}_{1i}-{\bf p}_{1f}={\bf p}_{2f}-{\bf p}_{2i}$, and define
\begin{mathletters}
\begin{equation}
   \Delta T\equiv T^{a_1f_1}_{nn}(S)-T^{a_1f_1}_{pp}(S)\ . \label{eqno4a}
\end{equation}
Then
\begin{equation}
   \Delta T = - g_{a_1N\!N}g_{f_1N\!N}
      \frac{\langle a_1^\circ|H_{em}|f_1\rangle}{({\bf q}^2 + m^2_A)^2}
      \left\{\mbox{\boldmath $\sigma$}_1\!\cdot\!
             \mbox{\boldmath $\sigma$}_2 +
      \frac{(\mbox{\boldmath $\sigma$}_1\!\cdot\!{\bf q})
            (\mbox{\boldmath $\sigma$}_2\!\cdot\!{\bf q})}{m^2_A}
      \left(2+\frac{{\bf q}^2}{m^2_A}\right)\right\}\ , \label{eqno4b}
\end{equation}
where we have dropped nonlocal terms and used the isospin convention
$\tau_3|p\rangle = +|p\rangle$.  Finally, we take the Fourier transform
to arrive at
\begin{equation}
   \Delta V = - \frac{2}{3}\frac{g_{a_1N\!N}g_{f_1N\!N}}{4\pi}
     \mbox{\boldmath $\sigma$}_1\!\cdot\!\mbox{\boldmath $\sigma$}_2
     \frac{\langle a_1^\circ|H_{em}|f_1\rangle}{2m_A}\,e^{-m_Ar}\ ,
                                      \label{eqno4c}
\end{equation}
\end{mathletters}
where $m_A=(m_{a_1}+m_{f_1})/2$. This is of the familiar form of the
dominant term of the
class III CSB potential due to $\rho$--$\omega$ mixing~\cite{footnote},
with the exception of the coefficient $-{\textstyle\frac{2}{3}}
\mbox{\boldmath $\sigma$}_1\!\cdot\!\mbox{\boldmath $\sigma$}_2$
which takes the value +2 in the
$^1S_0$ partial wave of interest in the few-nucleon systems.  Therefore,
in these systems
axial-vector meson mixing will have the same CSB effect as vector meson
mixing.  Notice that the coordinate space potential in Eq.~(\ref{eqno4c})
is exponential with a longer range than a Yukawa of the same mass.

\section{ESTIMATES FOR COUPLING CONSTANTS}
In numerical calculations with Eq.~(\ref{eqno4c}) we have used two
estimates of the needed coupling constants $g_{a_1N\!N}$ and
$g_{f_1N\!N}$. The first estimate comes from a model by Sudarshan for
strong, weak, and electromagnetic interactions~\cite{Sudarshan}.
The strong-interaction Lagrangian of this model has been used to study
violation of time-reversal invariance in low-energy $N\!N$ scattering due
to interference of vector and tensor $a_1N\!N$ couplings~\cite{Bryan}. The
axial-vector coupling constants of this model were found to be barely
consistent with these experimental tests of time-reversal asymmetry.
Sudarshan's model assumes that the $\rho$, $a_1$, and $f_1$ are coupled
to the nucleon and $\Delta(1232)$ in an $SU(4)$ symmetry scheme.
The predicted coupling constants are
$g_{a_1N\!N}=5g_{\rho N\!N}/(3\sqrt{2})$ and
$g_{f_1N\!N}=g_{\rho N\!N}/\sqrt{2}$. The ratio
$g_{a_1N\!N}/g_{f_1N\!N}=5/3$ is the same as that obtained from the
textbook derivation of $g_A/g_V=5/3$ using expectation values of
$I_3\sigma_z$ versus $\sigma_z$ between proton states with $SU(6)$
wave functions.
The extra factor of $1/\sqrt{2}$ is peculiar to Sudarshan's model
and arises from setting the (time-reversal asymmetrical) tensor coupling
of the $a_1$ equal to the direct coupling so that $g_{a_1N\!N}\approx 1.18
g_{\rho N\!N}$, a number desired at that time when the empirical $g_A/g_V$
was thought to be $\approx 1.18$. Taking the Sudarshan model literally,
we find that $g^2_{a_1N\!N}/4\pi \approx 2.8$ and
\begin{equation}
   \frac{g_{a_1N\!N}g_{f_1N\!N}}{4\pi} = \frac{5}{6}\,
   \frac{g^2_{\rho N\!N}}{4\pi} \approx 1.7\ ,    \label{eqno5}
\end{equation}
where the numerical values are derived from the vector dominance
hypothesis, universality ($g_{\rho N\!N} \approx g_{\rho\pi\pi} \approx
g_\rho$), and the data on $\Gamma(\rho \rightarrow e^+ e^-)$~\cite{PDG},
which gives $g^2_{\rho}/ 4\pi \approx 2.02$.

The second estimate of the needed coupling constants uses the idea of
axial-vector dominance to relate $g_{a_1N\!N}$ to the observed
axial-vector coupling constant $g_A(0)$. The assumption that the nucleon
matrix elements of the isovector axial-vector hadronic current,
\begin{equation}
   \langle N_{p'}|A^i_\mu| N_{p}\rangle
   =\bar{N}_{p'}{\textstyle\frac{1}{2}}\tau^i
     \left[g_A(q^2)\gamma_{\mu}\gamma_5 + h_A(q^2)q_\mu\gamma_5\right]
     N_p e^{i q\cdot x}\ ,              \label{eqno6}
\end{equation}
are dominated by
the contribution of the lowest-lying axial-vector mesons yields
immediately the relation
\begin{mathletters}
\begin{equation}
    g_A(0)=\frac{ f_{a_1} g_{a_1pp}}{m^2_{a_1}}\ ,  \label{eqno7a}
\end{equation}
where $g_A(0)=1.2573\pm 0.0028$~\cite{PDG}, and we define $g_{a_1pp}$
as in Eq.~(\ref{eqno1}). The decay constant
$f_{a_1}$ is defined by the isovector $a_1$-to-vacuum matrix element of
the hadronic axial-vector current
\begin{equation}
   \langle 0|A^i_\mu(0) |a^j_1\rangle = \delta^{ij}
      f_{a_1}\epsilon_\mu\ ,               \label{eqno7b}
\end{equation}
where $\epsilon_\mu$ is the polarization vector.
The definition (\ref{eqno7b}) is analogous to that of the pion decay
constant $f_\pi \approx 93$ MeV and the ``vector decay constant" $f_\rho
\equiv m^2_\rho/g_\rho$, and so it is a factor of $\sqrt{2}$ smaller than
the convention of Ref.~\cite{Tsai}.
The $a_1$ mass and decay constant have recently been calculated from
lattice QCD~\cite{Wingate}.  The simulation finds $m_{a_1} = 1250 \pm 80$
MeV and
\begin{equation}
     f_{a_1} = 0.21 \pm 0.02\,{\rm GeV}^2\ ,    \label{eqno7c}
\end{equation}
in our convention (\ref{eqno7b}).
The decay constant can be obtained from the measured partial width of
the decay $\tau^- \rightarrow a^-_1 + \nu_{\tau}$~\cite{Tsai,FR} but
the calculation is complicated by the large width of the $a_1$ and of
the intermediate $\rho$ mesons in the decay to three pions.
An empirical value of $f_{a_1}$ has been obtained in which
Breit-Wigner forms were taken for the $\rho$ and $a_1$ single-particle
contributions to vector and axial-vector spectral functions of
$\tau$ decay~\cite{Pham}.  Redoing this calculation with the
``new world average" branching fractions~\cite{Heltsley}
${\cal B}(\tau^- \rightarrow 3 \pi \nu_\tau)
= (17.73 \pm 0.28)\%$ and ${\cal B}(\tau^- \rightarrow  \rho^- \nu_\tau)
= (24.91 \pm 0.21\%)$, we obtain $f_{a_1}= 0.23\, {\rm  GeV}^2$, in good
agreement with the lattice result.
With the lattice value of $f_{a_1}$, Eq.~(\ref{eqno7a}) gives
\begin{equation}
    g_{a_1pp} \approx 9.4 \pm 1.2\ ,   \label{eqno7d}
\end{equation}
\end{mathletters}
for the $a_1$ mass in the range of $1250\pm 80$ MeV.

This second estimate of $g_{a_1pp}^2/4\pi\approx 7$ is larger than the
one from Sudarshan's model but still rather
conservative compared to other estimates in the literature.  It is
within two standard deviations of the
coupling constant $g_{a_1pp}^2/4\pi\approx 29\pm12$
(in our convention of Eq.~(\ref{eqno1}))
obtained from a best fit to the discrepancy functions of a forward
dispersion relation analysis of $N\!N$ data~\cite{Kroll}. In addition,
Eq.~(\ref{eqno7d}) is about a factor of two smaller than the $g_{a_1pp}$
predicted by chiral-invariant effective Lagrangians~\cite{effL} where
$ g_{a_1pp}/(2 m_{a_1}) = \sqrt{m^2_{a_1} -
m^2_{\rho}}f/(m_{\rho}m_{\pi})$ and $f$ is the
$\pi N\!N$ pseudovector coupling constant ($f^2/4\pi= 0.075$).
The disagreement worsens if one uses empirical masses instead of
assuming $m_{a_1}\approx \sqrt{2}m_\rho$ which is suggested by the
vector dominance argument of \cite{effL} but is not supported by
experiment. The latter chiral relationship of \cite{effL} has been used
to discuss the role of the $a_1$ meson in the $N\!N$
interaction~\cite{Durso}, and to construct two-meson-exchange
contributions to the $N\!N$ interaction when one or both nucleons
contains a meson pair vertex~\cite{Stoks}. On the other hand,
Eq.~(\ref{eqno7d}) is near what one would expect ($g_{a_1 N\!N}\approx
\sqrt{2}g_{\rho N\!N}$) from dynamical generation of the vector and
axial-vector gauge fields ($\rho$ and $a_1$) from a meson-quark
Lagrangian~\cite{Del}.  The latter theory predicts the relation
$m_{a_1}\approx \sqrt{3}m_\rho \approx 1330$ MeV, close to experiment
and a motivation for considering a dependence of our results upon the
$a_1$ mass.

In addition, we need the strength of the coupling of the $f_1(1285)$
meson to the nucleons. It is estimated in the same way as used
previously for the pseudoscalar meson couplings~\cite{CSeta}.
The axial-vector mesons $f_1(1285)$ and $f_1(1420)$
are mixtures of nonstrange ($NS$) and strange ($S$) quark states
$|NS\rangle = |\bar{u}u + \bar{d}d\rangle/\sqrt{2}$ and
$|S\rangle = |\bar{s}s\rangle$ such that the standard mixing equations
hold:
\begin{mathletters}
\begin{eqnarray}
   |f_1(1285)\rangle &=& \cos\phi_A|A_{NS}\rangle -
                         \sin\phi_A|A_{S}\rangle\ , \label{eqno8a}\\
   |f_1(1420)\rangle &=& \sin\phi_A|A_{NS}\rangle +
                         \cos\phi_A|A_{S}\rangle\ . \label{eqno8b}
\end{eqnarray}
\end{mathletters}
The small axial-vector mixing angle $\phi_A$ in this basis is directly
estimated from the
observed~\cite{PDG} widths of the axial-vector mesons $f_1(1285)$ and
$f_1(1420)$.  An update of the estimate of Ref.~\cite{AS} finds
$\phi_A\approx 12^\circ$. These mixing equations in the $SU(3)$
octet-singlet basis,
$|8\rangle = |\bar{u}u + \bar{d}d - 2\bar{s}s \rangle /\sqrt{6}$ and
$|0\rangle = |\bar{u}u + \bar{d}d +  \bar{s}s \rangle /\sqrt{3}$, take
the form
\begin{mathletters}
\begin{eqnarray}
   |f_1(1285)\rangle &=& \cos\theta_A|A_{8}\rangle -
                         \sin\theta_A|A_{0}\rangle\ , \label{eqno9a}\\
   |f_1(1420)\rangle &=& \sin\theta_A|A_{8}\rangle +
                         \cos\theta_A|A_{0}\rangle\ , \label{eqno9b}
\end{eqnarray}
\end{mathletters}
where $\theta_A = \phi_A - \arctan \sqrt{2}$.
Now the needed coupling $g_{f_1(1285)pp}$ is
estimated with the $SU(3)$ structure
constants~\cite{FR} and the quark model assumption that the strange quark
state $|S\rangle = (-\sqrt{2} |8\rangle +  |0\rangle)/\sqrt{3}$ does
not couple to the nucleon.  The latter Zweig rule assumption fixes the
coupling of the singlet in broken $SU(3)$ to be $g_{A_0 pp} = \sqrt{2}
g_{A_8 pp}$.  Within the axial-vector octet the $SU(3)$ invariant
axial-vector--baryon--baryon (ABB) couplings include
$g_{A_8 pp} = + g(3f - d)/\sqrt{3}$,
where the $f$ and $d$ are the strengths of the
antisymmetric and symmetric ABB couplings. They are normalized, $f+d=1$,
so that $g$ is $g_{a_1pp}$.  Inserting the assumed singlet coupling
into the octet-singlet mixing relations (9) and re-expressing the result
in the $NS-S$ basis of (8), one finds the simple result
\begin{mathletters}
\begin{eqnarray}
    g_{f_1(1285)pp} &=& \cos\phi_A (3f-d) g_{a_1pp}\ , \label{eqno10a}\\
    g_{f_1(1420)pp} &=& \sin\phi_A (3f-d) g_{a_1pp}\ . \label{eqno10b}
\end{eqnarray}
\end{mathletters}
The derived ratio
\begin{equation}
    g_{f_1(1420)pp}/ g_{f_1(1285)pp} = \tan\phi_A \approx 0.21\ ,
                                    \label{eqno11}
\end{equation}
is clearly seen in the $NS-S$ basis of (8) if $g_{A_Spp} = 0$ (Zweig
rule).
The axial-vector current $f/d$ ratio, obtained from the Cabibbo theory
of semileptonic decays of baryons, is now known to be
$0.575 \pm 0.0165$~\cite{Close}. Continuing in the spirit of axial-vector
dominance, we
assume the structure constants for the coupling of the axial-vector
mesons to the nucleons to have the same ratio and find,
with the small mixing angle $\phi_A\approx 12^\circ$,
\begin{equation}
    g_{f_1pp}\approx (0.45 \pm 0.025)g_{a_1pp}\ .  \label{eqno12}
\end{equation}
We neglect mixing of the $a_1$ with the $f_1(1420)$ as it is a factor of
five smaller than (\ref{eqno12}) from (\ref{eqno11}).

Next we examine the assumptions made to arrive at (\ref{eqno12}).
The first was that the $f_1(1420)$ does belong to the axial $\bar{q}q$
nonet of Eqs.~(8) and (9). This assumption has been supported~\cite{Aihara}
and attacked~\cite{Bityukov} on experimental grounds and remains
uncertain~\cite{PDG}. On the theory side, a more elaborate mixing scheme
with nonstrange and strange quark states and gluon states has suggested
specific admixtures of the {\em three} isoscalar axial-vector mesons
$f_1(1285)$, $f_1(1420)$, and $f_1(1510)$~\cite{Fritzsch}.  In this
scheme the $f_1(1285)$ is primarily a $NS$ $\bar{q} q$ state with a
coefficient of about 0.9 and the other two have small $NS$ components,
so that the estimate (\ref{eqno12}) from the nonet picture of Eq.~(8)
would not be altered by more than 10\%.
The second (Zweig rule) assumption about the singlet coupling
strength ($g_{0pp} = \sqrt{2} g_{8pp}$) can be compared to the
pseudoscalar coupling constants of the most recent Nijmegen potential
model of the nucleon-nucleon interaction~\cite{Nijmegen}.  The fitting
procedure of the one-boson-exchange
potential Nijm93 fixes the $\pi^0 N\!N$ coupling constant, $f/d$, and the
mixing angle $\theta_P$.  A search on the $N\!N$ data finds
$g_{0pp} = 1.9 g_{8pp}$ at the meson poles, somewhat
larger than the Zweig rule assumption.

Finally, taking central values from Eqs.~(\ref{eqno7a}), (\ref{eqno7c}),
and (\ref{eqno12}) we find
\begin{equation}
   \frac{g_{a_1pp}g_{f_1pp}}{4\pi}\approx
      0.45 \frac{m^4_{a_1}(g_A(0))^2} {4\pi\,f^2_{a_1}}\approx 3.1\ ,
                                    \label{eqno13}
\end{equation}
for an assumed $a_1$ mass of 1250 MeV.  Note that the CSB potential
Eq.~(\ref{eqno4c}) depends on three powers of the poorly known $a_1$
mass if the coupling constants of Eqs.~(7) and (\ref{eqno12}) are used.
We do not assume charge symmetry breaking in the meson-nucleon-nucleon
vertices, as has been suggested for vector meson exchange, first with
explicit vertex models~\cite{Gardner},
and then discussed in more general terms~\cite{Cohen}. Charge symmetry
breaking in our preferred model (7) for the $AN\!N$ vertices could arise
from a difference in $g_A(0)_p$ and $g_A(0)_n$.  This difference has
recently been calculated within the external field QCD sum rule
approach and found to be at the 1\% level or less~\cite{Jin}, and we
disregard it.

\section{NUMERICAL RESULTS}
Now we turn to the charge symmetry breaking in the $N\!N$ interaction from
the model potential (\ref{eqno4c}) with coupling constants given in
Eqs.~(\ref{eqno5}) or (\ref{eqno13}).
Traditional measures of nuclear charge asymmetry have been obtained from
the positive value for the difference
$\Delta a = |a_{nn}| - |a_{pp}| \approx {\cal O}$(1 fm) of the
$N\!N$ singlet scattering lengths
and the positive value for the ${^3}$H--${^3}$He binding-energy
difference $\Delta E \approx {\cal O}$(100 keV).
The measures~\cite{CN95}
\begin{eqnarray*}
   \Delta a_{\rm exp} & \equiv & (|a_{nn}| - |a_{pp}|)
          \approx  +1.1 \pm 0.6\;{\rm fm}\ , \\
   \Delta r_{0\;\rm exp} & \equiv & (r_{nn} - r_{pp})
          \approx -0.02 \pm 0.11\;{\rm fm}\ , \\
   \Delta E_{\rm exp} & \equiv & ({^3}{\rm H}-{^3}{\rm He})
          \approx 76 \pm 24 \;{\rm keV}\ ,
\end{eqnarray*}
are quoted after correction of experiment for direct
electromagnetic effects and are quite consistent in sign and magnitude.
A positive $\Delta a$ reflects an interaction between two neutrons which
is more attractive than between two protons and more binding energy is
provided for ${^3}$H as compared to ${^3}$He. The binding-energy difference
$\Delta E$ is quantitatively tied to the $^1S_0$ effective range
parameters $\Delta a$ and $\Delta r_0$, as has been known for a long
time, and demonstrated in recent three-body calculations
(see Ref.~\cite{CN95} for a discussion).

The theoretical shifts in $a$
and $r$ are obtained by adding the model for $\Delta V = V_{nn} - V_{pp}$
to a model for the charge symmetric reaction.  For the latter we choose
two new
Nijmegen potential models~\cite{Nijmegen} which give an excellent
description of the data and an older potential~\cite{dtRS} which has
a ``super-soft
core" and should therefore allow the largest effects of the short-range
model (4). Because all three charge symmetric potentials are repulsive
at short range, we did not include form factors in the CSB model (4).
The results are displayed in Table I which demonstrates that
the dependence of the results upon the charge symmetric potential are
noticeable but the CSB effect from axial-vector
meson mixing is rather small.  The mass of the $a_1$ was taken to be
1250 MeV in Table I; an uncertainty of 80 MeV in the $a_1$ mass in the
couplings of Eq.~(\ref{eqno13}) gives an uncertainty of 0.02 fm in the
calculated $\Delta a$.
We also show two estimates of $\Delta E$,
the first $\Delta E_{\rm GS}$ obtained from a relationship shown
in Ref.~\cite{CN95} from the triton calculations of Ref.~\cite{GS}, and
the second $\Delta E_{\rm FF}$ obtained from empirical charge form
factors of ${^3}$H and ${^3}$He with the aid of the hyperspherical
formula~\cite{CB,BCS}.  Both calculations are expected to overestimate
$\Delta E$ somewhat, see Ref.~\cite{CN95} for details.

The results of Table I can be compared with the empirical measures above
and with the effect of $\pi$--$\eta$--$\eta^\prime$ mixing
($\Delta a \approx +0.26$ fm from Ref.~\cite{CSeta}) and of
$\rho$--$\omega$ mixing ($\Delta a \approx +1.5$ fm from Ref.~\cite{CB}).
Another source of CSB
suggested is two-pion exchange ($\Delta a \approx +1$ fm
from Ref.~\cite{CN95}).  Simultaneous $\gamma$--$\pi$ exchange does not
contribute to CSB to the lowest order~\cite{CF}.
By all these measures, the contribution of axial-vector mixing to
nuclear charge asymmetry is small but in the direction of
experiment.

We conclude by returning to the phenomenological Lagrangian analysis of
isospin breaking~\cite{vkfg,Bira} which gives a more general measure of
charge symmetry breaking. First we compare the strength of our
meson-mixing potential with the expected strength $(\gamma_{\sigma} \sim
\epsilon m^2_{\pi}/f^2_{\pi}\Lambda^2)$ of the $N\!N$ contact
force $\gamma_{\sigma}(\bar{N}\frac{\tau_3}{2}\mbox{\boldmath
$\sigma$}N)\!\cdot\!(\bar{N}\mbox{\boldmath $\sigma$}N)$.
The momentum space form of Eq.~(\ref{eqno4b})
becomes in the small ${\bf q}^2$ limit (${\bf q}^2 \ll m^2_A$)
\begin{equation}
 V({\bf q})   = + g_{a_1N\!N}g_{f_1N\!N}
      \frac{\langle a_1^\circ|H_{em}|f_1\rangle}{2 m^4_A}
            (\mbox{\boldmath $\sigma$}_1\!\cdot\!
             \mbox{\boldmath $\sigma$}_2)
      \left(\frac{\tau_3(1)}{2} + \frac{\tau_3(2)}{2}\right)
, \label{eqno14}
\end{equation}
where we have explicitly reinstated the isospin operators of
Eq.~(\ref{eqno2}) to facilitate the comparison.  Then the axial-vector
mixing mechanism of this paper gives a contribution  to $\gamma_{\sigma}$
of
\begin{equation}
 \gamma_{\sigma}^{a_1f_1} = g_{a_1N\!N}g_{f_1N\!N}
      \frac{\langle a_1^\circ|H_{em}|f_1\rangle}{2 m^4_A}
       \equiv c_{a_1f_1}(\epsilon m^2_{\pi}/f^2_{\pi}\Lambda^2).
 \label{eqno15}
\end{equation}
Inserting  Eq.~(\ref{eqno7d}), Eq.~(\ref{eqno12}), and the assumed
universal value of the mixing matrix element
$\langle a_1^\circ|H_{em}|f_{1\,NS}\rangle \approx -0.005$ GeV$^2
= -0.85( \epsilon m^2_{\pi})$ into Eq.~(\ref{eqno15}) and choosing the
large-mass scale $\Lambda = m_{\rho}$, determines
$c_{a_1f_1} \simeq -0.03$, a number very far from ${\cal O}$(1). There
is a reduction of the dimensionless coefficient from that of vector-meson
mixing by about a factor of three since $m^2_{a_1}\sim 3 m^2_{\rho}$.
Such a reduction will always occur when the ``object" exchanged and
mixed in the  $t$-channel
is more massive than $\Lambda$ but it alone cannot account for an
unnatural coefficient in such stark contrast with the vector and
pseudoscalar mixing cases.

It is instructive to breakdown this unnatural result into a comparison
of coupling constants with the $\rho$--$\omega$ mixing mechanism. This is
because the (on-mass-shell) mixing matrix element is the same as that
of the latter which has been found natural ${\cal O}(\epsilon m^2_{\pi}$)
in Ref.~\cite{vkfg} (which we follow by parameterizing isospin violation
by $\epsilon \sim 0.3$). The coupling constant $g_x$ of this zero-range
Lagrangian is natural if $g_x/m_x \sim 1/f_{\pi}$~\cite{vkfg,Friar}.
Thus $g_{\rho N\!N}/m_{\rho} \simeq 0.60/f_{\pi}$ is nearly the same as
$g_{a_1N\!N}/m_{A} \simeq 0.70/f_{\pi}$ from Eq.~(\ref{eqno7d}) and they
are both natural.  Even the twice as large axial-vector coupling
constants of Refs.~\cite{Kroll,effL} are natural in that the
dimensionless number is {\em near} unity. The difference between the
natural potential of $\rho$--$\omega$ mixing and the unnatural one of
$a_{1}$--$f_{1}$ mixing lies in the contrast between the natural
coupling of the $I=0$ vector meson $g_{\omega N\!N}/m_{\omega} \sim
3g_{\rho N\!N}/m_{\rho} \simeq 1.8/f_{\pi}$ and the unnatural coupling
of the $I=0$ axial-vector $g_{f_1N\!N}/m_{A} \sim 0.5 g_{a_1N\!N}/m_{A}
\simeq 0.3/f_{\pi}$ from Eq.~(\ref{eqno12}). This factor of six
reduction can in turn be traced to the  broken $SU(3)$ analysis of
vector and axial-vector couplings to nucleons.  A treatment analogous to
that of Eqs.~(8)--(\ref{eqno12}) can be made for the vector meson
coupling constants. With the aid of the vector dominance hypothesis, one
finds i) a very small mixing angle $\phi_V$ (``ideal mixing") and ii)
pure F vector-baryon-baryon coupling so that $f=1$ and
$d=0$~\cite{CMcN}.  Thus from Eq. (10a), the small coupling $g_{f_1N\!N}$
is due to the ratio $d/f = 1.74 \pm 0.05$ (chosen from the axial-vector
current $d/f$) rather than the $d/f = 0$ of the vector mesons. This
guess for the coupling constant $g_{f_1N\!N}$ seems reasonable and we
must leave our result at that.  In any event, the reduction by a factor
of about 18 from the $\rho$--$\omega$ mixing case has been exposed.

In conclusion, the contribution of axial-vector mixing to nuclear charge
asymmetry in two few-body mirror systems is small but in the direction
of experiment.  The dimensionless coefficient $c_{a_1f_1}$ which
characterizes the charge asymmetric potential is much smaller that
unity, indicating that axial-vector mixing (with the values for coupling
constants and matrix elements used here) does not saturate the
corresponding term of the phenomenological Lagrangian~\cite{vkfg,Bira}.
If this conclusion remains true as more is learned about the axial-vector
mesons, one might suggest that nonresonant $\rho$--$\pi$ exchange between
two nucleons should be examined.  This suggestion is based upon a
calculation in the same nuclear systems~\cite{CN95} which indicates that
nonresonant $\pi$--$\pi$ exchange might dominate the other short range
coefficient $\gamma_s$, if a convincing case for a suppression from its
on-shell value of the $\rho$--$\omega$ mixing element can be made.

\acknowledgments
One of us (SAC) wishes to acknowledge helpful discussions with M.\ D.\
Scadron at the beginning of this project.   SAC is grateful for the
hospitality of TRIUMF and the University of Melbourne and acknowledges
support in part by NSF grant PHY-9408137.  We appreciate a
correspondence with Peter Kroll on coupling constants, and a very
helpful remark by Jim Friar.

\begin{table}
\caption{The axial-vector ($a_{1}$--$f_{1}$) mixing contribution to
         $\Delta a$ and $\Delta r_0$, and to $\Delta E$, the binding-energy
         difference between $^{3}$He and $^{3}$H. The charge-asymmetric
         potentials are distinguished by the two choices
         [Eqs.~(\protect\ref{eqno5}) or (\protect\ref{eqno13})]
         of axial-vector couplings to the nucleon.
         The Nijmegen Reid-like (Reid93), one-boson exchange (Nijm93),
         and de Tourreil-Rouben-Sprung (dTRS) potentials are the
         charge-symmetric potentials $V(CS)$ used in the calculation of
         $\Delta a$, $\Delta r_0$, and the estimate of $\Delta E_{\rm GS}$
         based on these effective-range parameters.
         Another estimate, $\Delta E_{\rm FF}$, is obtained from the
         ``model independent" hyperspherical formula.}
\begin{tabular}{cccccc}
   $V(CS)$ & $g_{a_{1}N\!N}g_{f_{1}N\!N}/4\pi$
         & $\Delta a$ (fm) & $\Delta r_0$ (fm)
         & $\Delta E_{\rm GS}$ (keV) & $\Delta E_{\rm FF}$ (keV) \\
\tableline
 Reid93 & 1.7 & +0.06 & --0.001 &  +4 &   +8  \\
 Nijm93 & 1.7 & +0.07 & --0.001 &  +4 &   +8  \\
 dTRS   & 1.7 & +0.09 & --0.002 &  +7 &   +8  \\[2mm]
 Reid93 & 3.1 & +0.11 & --0.002 &  +8 &  +15  \\
 Nijm93 & 3.1 & +0.13 & --0.003 & +10 &  +15  \\
 dTRS   & 3.1 & +0.16 & --0.004 & +13 &  +15
\end{tabular}
\label{table1}
\end{table}


\begin{thebibliography}{99}

\bibitem{vkfg} U. van Kolck, J. L. Friar, and T. Goldman,
Phys. Lett. {\bf  B371} (1996) 169.

\bibitem{Bira} U. van Kolck, in {\em Low Energy Effective Theories and
QCD}, D.-P. Min, ed. (Man Lim Won, Seoul, 1995);
Thesis, University of Texas (1993).

\bibitem{Weinberg} S. Weinberg, Physica {\bf 96A} (1979) 327; Phys.
Lett. {\bf B251} (1990) 288.

\bibitem{GM} A. Manohar and H. Georgi, Nucl. Phys. {\bf B234} (1984) 189;
H. Georgi, Ann. Rev. Nucl. Part. Sci. {\bf 43}  (1993) 209.

\bibitem{PSC} P. C. McNamee, M. D. Scadron,
and S. A. Coon, Nucl. Phys. {\bf A249} (1975) 483.

\bibitem{PSC77} S. A. Coon, M. D. Scadron, and P. C. McNamee, Nucl.
Phys. {\bf A287} (1977) 381.

\bibitem{CSeta} S. A. Coon and M. D. Scadron, Phys. Rev. {\bf C26} (1982) 562.

\bibitem{CB} S. A. Coon and R. C. Barrett, Phys. Rev. {\bf C36} (1987) 2189.

\bibitem{vanoers} G. A. Miller and W. T. H. van Oers, in {\em Symmetries
and Fundamental Interactions in Nuclei}, W. Haxton and E. M. Henley,
eds. (World Scientific, Singapore, 1995).

\bibitem{CS95} S. A. Coon and M. D. Scadron, Phys. Rev. {\bf C51}
(1995) 2923.

\bibitem{CG} S. Coleman and S. Glashow, Phys. Rev. {\bf 134} (1964) B671.

\bibitem{Coleman} S. Coleman, {\it Aspects of Symmetry} (Cambridge
University Press, Cambridge, 1985) pp. 23-35.

\bibitem{Ecker} G. Ecker, J. Gasser, A. Pich, and E. de Rafael, Nucl.
Phys. {\bf B321} (1989) 311; J. F. Donoghue, C. Ramirez, and G.
Valencia, Phys. Rev. {\bf D39} (1989) 1947.

\bibitem{OZ1} H. B. O'Connell, B. C. Pearce, A. W. Thomas, and A. G.
Williams, Phys. Lett. {\bf B336} (1994) 1.

\bibitem{qdep} T. Goldman, J. A. Henderson, and A. W. Thomas, Few-Body
Systems {\bf 12} (1992) 123; T. Hatsuda, E. M. Henley, Th. Meissner,
and G. Krein, Phys. Rev. {\bf C49} (1994) 452; K. Maltman, Phys. Lett.
{\bf B313} (1993) 203.

\bibitem{BD} The convention for gamma matrices is that of J. D. Bjorken
and S. D. Drell, {\em Relativistic Quantum Mechanics} (McGraw Hill,
NY, 1964).

\bibitem{footnote} We define the $AN\!N$ coupling with a factor of 1/2
which arises naturally from an identification of the axial-vector mesons
and the axial-vector current.  A similar factor appears in the $V\!N\!N$
couplings of Refs.~\cite{PSC,CB}, so that vector meson universality takes
the form $g_{\rho NN} \approx g_{\rho\pi\pi} \approx g_\rho$.

\bibitem{footnote2} The class IV CSB potential arising from vector meson
mixing is due entirely to the tensor couplings~\cite{vanoers,CS95}. By
neglecting the tensor couplings of the axial-vector mesons, we do not
attempt to evaluate a class IV CSB potential from this source.

\bibitem{PDG} Particle Data Group (PDG), L. Montanet et al.,
Phys. Rev. {\bf D50}, Part I (1994) 1173.

\bibitem{Sudarshan} E. C. G. Sudarshan, Proc. Roy. Soc. {\bf A305}
(1968) 319.

\bibitem{Bryan} R. Bryan and A. Gersten, Phys. Rev. Lett. {\bf 26}
(1971) 1000; J. Binstock, R. Bryan, and A. Gersten, Ann. Phys. (N.Y.)
{\bf 133} (1981) 355.

\bibitem{Tsai} Y.-S. Tsai, Phys. Rev. {\bf D4} (1971) 2821.

\bibitem{Wingate} M. Wingate, T. DeGrand, S. Collins, and U. M.
Heller, Phys. Rev. Lett. {\bf 74} (1995) 4596.

\bibitem{FR} Fayyazuddin and Riazuddin, {\em A Modern Introduction to
Particle Physics} (World Scientific, Singapore, 1992)

\bibitem{Pham} T. N. Pham, Phys. Rev. {\bf D46} (1992) 2976; see
equation 33.

\bibitem{Heltsley} B. K. Heltsley, Nucl. Phys. B (Proc. Suppl.)
{\bf 40} (1995) 413; see table 5.

\bibitem{Kroll} W. Grein and P. Kroll, Nucl. Phys. {\bf A338}
(1980) 332; P. Kroll, $\pi N$ Newsletter, no. 2, May 1990.  We believe
that the vector and axial-vector coupling constants of these papers
correspond to those of an effective Hamiltonian without the factor of
1/2 appearing in Eq.~(\ref{eqno1}) and discussed in Ref.~\cite{footnote}.

\bibitem{effL} S. Weinberg, Phys. Rev. Lett. {\bf 18} (1967) 188; J.
Schwinger, Phys. Lett. {\bf 24B} (1967) 473; J. Wess and B. Zumino,
Phys. Rev. {\bf 163} (1967) 1727.

\bibitem{Durso} J. W. Durso, G. E. Brown, and M. Saarela, Nucl. Phys.
{\bf A430} (1984) 653.

\bibitem{Stoks} Th. A. Rijken and V. G. J. Stoks, submitted to Phys.
Rev. C.

\bibitem{Del} R. Delbourgo and M. D. Scadron, Mod. Phys. Lett.
{\bf A10} (1995) 251.

\bibitem{AS} M. Anselmino and M. D. Scadron, Nuovo Cim. {\bf 104A}
(1991) 1091.

\bibitem{Close} F. E. Close and R. G. Roberts, Phys. Lett.
{\bf B316} (1993) 165.

\bibitem{Aihara} H. Aihara et al., Phys. Rev. {\bf D38} (1988) 1.

\bibitem{Bityukov} S. I. Bityukov et al., Phys. Lett. {\bf B203}
(1988) 327.

\bibitem{Fritzsch} M. Birkel and H. Fritzsch, Phys. Rev. {bf\ D53}
 (1996)6195.

\bibitem{Nijmegen} V. G. J. Stoks, R. A. M. Klomp, C. P. F. Terheggen,
and J. J. de Swart, Phys. Rev. {\bf C49} (1994) 2950.

\bibitem{Gardner} S. Gardiner, C. J. Horowitz, and J. Piekarewicz,
Phys. Rev. Lett. {\bf 75} (1995) 2462.

\bibitem{Cohen} T. D. Cohen and G. A. Miller, Phys. Rev.
{\bf C52} (1995) 3428.

\bibitem{Jin} X. Jin, hep-ph/9602298.

\bibitem{CN95} S. A. Coon and J. A. Niskanen, Phys. Rev.
{\bf C53} (1996) 1154.

\bibitem{dtRS} R. de Tourreil, B. Rouben, and D. W. L. Sprung,
Nucl. Phys. {\bf A242} (1975) 445.

\bibitem{GS} B. F. Gibson and G. J. Stephenson, Jr., Phys.
Rev. {\bf C8} (1973) 1222.

\bibitem{BCS} R. A. Brandenburg, S. A. Coon, and P. U. Sauer,
Nucl. Phys. {\bf A294} (1978) 305. See Ref.~\cite{CB} for discussion of
updates to data in this method.  The data on the $^3$He and $^3$H charge
form factors are from A. Amroun et al., Nucl. Phys. {\bf A579} (1994) 596.

\bibitem{CF} J. L. Friar and S. A. Coon, Phys. Rev.
{\bf C53} (1996) 588.

\bibitem{Friar} J. L. Friar, to appear in Few-Body Systems Supplement
Series.

\bibitem{CMcN}  See, for example, S. A. Coon and P. C. McNamee, Nucl.
Phys. {\bf A322} (1979) 267.
\end{thebibliography}
\end{document}